\documentclass[aps,prl,floatfix,nofootinbib,showpacs,twocolumn,superscriptaddress]{revtex4-1}

\usepackage{mathrsfs}
\usepackage{amsmath}
\usepackage{amssymb}
\usepackage{bbold}
\usepackage{color}
\usepackage{graphicx}
\usepackage{soul}

\usepackage[mathscr,scaled=1.15]{urwchancal}
\DeclareFontFamily{OT1}{pzc}{}
\DeclareFontShape{OT1}{pzc}{m}{it}%
{<-> s * [1.15] pzcmi7t}{}
\DeclareMathAlphabet{\mathpzc}{OT1}{pzc}{m}{it}

\definecolor{purple}{rgb}{0.5,0,0.5}
\definecolor{blue}{rgb}{0.0,0,0.9}
\definecolor{prdblue}{rgb}{0.133,0.118,0.498}
\usepackage[colorlinks=true, pdfstartview=FitV, linkcolor=prdblue, citecolor= prdblue, urlcolor=prdblue]{hyperref}

\begin{document}

\title{Exposing strangeness: projections for kaon electromagnetic form factors}

\author{Fei Gao}
\email{hiei@pku.edu.cn}
\affiliation{Department of Physics and State Key Laboratory of Nuclear Physics and Technology, Peking University, Beijing 100871, China}
\affiliation{Collaborative Innovation Center of Quantum Matter, Beijing 100871, China}

\author{Lei Chang}
\email{leichang@nankai.edu.cn}
\affiliation{School of Physics, Nankai University, Tianjin 300071, China}

\author{Yu-Xin Liu}
\email{yxliu@pku.edu.cn}
\affiliation{Department of Physics and State Key Laboratory of Nuclear Physics and Technology, Peking University, Beijing 100871, China}
\affiliation{Collaborative Innovation Center of Quantum Matter, Beijing 100871, China}
\affiliation{Center for High Energy Physics, Peking University, Beijing 100871, China}

\author{Craig~D.~Roberts}
\email{cdroberts@anl.gov}
\affiliation{Physics Division, Argonne National Laboratory, Argonne
Illinois 60439, USA}

\author{Peter~C.~Tandy}
\email{tandy@kent.edu}
\affiliation{Center for Nuclear Research, Department of
Physics, Kent State University, Kent, Ohio 44242, USA}

\date{13 March 2017}
%\date{22 February 2017}
%\date{04 January 2017}
%\date{29 July 2016}

\begin{abstract}
A continuum approach to the kaon and pion bound-state problems is used to reveal their electromagnetic structure.  For both systems, when used with parton distribution amplitudes appropriate to the scale of the experiment, Standard Model hard-scattering formulae are accurate to within 25\% at momentum transfers $Q^2 \approx 8\,$GeV$^2$.  There are measurable differences between the distribution of strange and normal matter within the kaons, \emph{e.g}.\ the ratio of their separate contributions reaches a peak value of $1.5$ at $Q^2 \approx 6\,$GeV$^2$.  Its subsequent $Q^2$-evolution is accurately described by the hard scattering formulae.  Projections for kaon and pion form factors at timelike momenta beyond the resonance region are also presented.  These results and projections should prove useful in planning next-generation experiments.
\end{abstract}

%%\keywords{
%%Abelian anomaly \and
%%continuum QCD \and
%%dynamical chiral symmetry breaking \and
%%$\pi$-meson elastic and transition form factors \and
%%parton distribution amplitudes}

\smallskip

\maketitle

%% Feldmann & Kroll http://arxiv.org/pdf/hep-ph/9709203.pdf

\noindent\emph{I.$\;$Introduction}.\,---\,%
%\section{Introduction}
%
%The concept of strangeness in particle physics is more than sixty years old \cite{GellMann:1953zza, Nishijima01031955}, predating but ultimately yielding the notion of a strange quark ($s$), partnered with the up- and down-quarks ($u$, $d$) which define the valence content of normal matter (protons, pions, etc.) \cite{GellMann:1964nj, Zweig:1981pd}.  Strange particles, kaons amongst them, were those hadrons with unexpectedly long lifetimes, whose decays are now understood to proceed via weak $s\to u$ transitions within the Standard Model of Particle Physics.  In fact, kaons, originally numbered amongst the ``V-particles'', were first seen in 1947 \cite{Rochester:1947mi}; and since then they have proved invaluable in the continuing development of the Standard Model (SM) paradigm.
%
Kaons and strange quarks are a bridge between strong- and electroweak-interactions.  For instance, they opened the first window on $CP$-violation, responsible for the matter-antimatter asymmetry in our Universe, and characterise strong interactions in a sector where the Higgs-generated quark current-mass cannot be treated as a perturbation, \emph{i.e}.\ a domain where flavour-dependence of the strong interaction becomes important and measurable.  Indeed, at perturbative Standard-Model scales the strange-to-up$+$down quark mass ratio is $2 m_s/[m_u+m_d]=27.3(7)$ \cite{Olive:2016xmw}; but, as the resolving scale is reduced, this ratio evolves so that, in the far infrared, its value is much smaller ($\sim 1.2\,$-$\,1.5$ \cite{Bhagwat:2003vw, Bowman:2005vx, Bhagwat:2006tu}) owing to emergent phenomena peculiar to the strong interaction.  Consequently, comparisons between kaon and pion properties provide direct access to the interplay between strong and electroweak mass-generating mechanisms.  Such qualities make the kaon a tantalising subject for study, providing a challenging target for both experiment and theory, and demanding that connections be drawn between them.

There are four types of kaon: $K^\pm$, $K^0$, $\bar K^0$, whose valence-quark content is, respectively, $u\bar s$, $\bar u s$, $d\bar s$, $\bar d s$, where $\bar q$ identifies an antiquark; and, within the SM's strong-interaction sector, quantum chromodynamics (QCD), one of the most pressing empirical challenges is to map the distribution of electric charge within the kaons.  Since the charge of $u$- and $s$-quarks is different, this translates into a fairly direct measure of the relative distribution of normal- and strange-matter within the kaon; and also, importantly, its scale dependence.  Further, given that the charge-conjugation operation executes $K^+ \leftrightarrow K^-$, $K^0 \leftrightarrow \bar K^0$, and assuming isospin symmetry (no difference between $u$- and $d$-quarks, other than their electric charge), there are just two distinct charge distributions: $u$ in $K^+$ is the same as $d$ in $K^0$; $\bar s$ in $K^+$ is the same as $\bar s$ in $K^0$; and these distributions also describe those in the charge-conjugated states.  A third distribution is accessible in the isospin-symmetric limit, \emph{viz}.\ the $u$ distribution in the pion, which is the elastic pion form factor itself, and this provides an excellent counterpoint.

At low momentum transfers, $Q^2 \lesssim 0.2\,$GeV$^2$, charged-pion and -kaon elastic form factors, $F_M(Q^2)$, $M=\pi^+$, $K^+$, can be measured directly by scattering high-energy mesons from atomic electrons \cite{Dally:1981ur, Dally:1982zk, Amendolia:1986wj, Amendolia:1984nz, Dally:1980dj, Amendolia:1986ui}.  These data constrain the charge radii: $r_\pi=0.657(12)\,$fm, $r_K=0.58(6)\,$fm.  The kaon is expected to be smaller because it contains the heavier $s$-quark \cite{Tarrach:1979ta, Ji:1990rd, Cardarelli:1994ix, Burden:1995ve, Maris:2000sk, daSilva:2012gf, Chen:2012txa, Chen:2016sno}.  Owing to kinematic limitations on the energy of meson beams and unfavorable momentum transfers, one must use other methods to reach higher spacelike $Q^2$.  Meson electroproduction off nucleon targets is a reliable tool \cite{Qin:2017lcd}, which has already been used for the pion out to $Q^2=2.45\,$GeV$^2$ \cite{Volmer:2000ek, Horn:2006tm, Tadevosyan:2007yd, Horn:2007ug, Blok:2008jy}.  Importantly, approved pion experiments \cite{E1206101, E1207105} will extend this reach to $Q^2\approx 8.5\,$GeV$^2$, \emph{i.e}.\ a domain upon which longstanding issues in QCD might be resolved \cite{Chang:2013nia}; and a forthcoming kaon experiment \cite{E12-09-011} can potentially provide kaon data out to $Q^2 \approx 5.5\,$GeV$^2$ \cite{Horn:2016rip}.  Existing and anticipated spacelike data are complemented by measurements of $e^+ e^-$ annihilation into $\pi^+ \pi^-$, $K^+ K^-$, which afford access to pion and kaon form factors at timelike momenta out to $t\approx 17\,$GeV$^2$ \cite{Seth:2012nn, Seth:2013eaa}.

The impetus for measuring $F_M(Q^2)$ at large momentum transfers is a need to understand and validate a strict prediction of QCD \cite{Lepage:1979zb, Lepage:1980fj, Efremov:1979qk}, \emph{viz}.\ $\exists Q_0 \gg \Lambda_{\rm QCD}$ such that
\begin{equation}
Q^2 F_M(Q^2) \stackrel{Q^2 > Q_0^2}{\approx} 16 \pi \alpha_s(Q^2) f_M^2 \mathpzc{w}_M^2(Q^2)\,,
\label{kaonUV}
\end{equation}
where $\alpha_s$ is the one-loop strong running coupling,
$\Lambda_{\rm QCD}\approx 0.3\,$GeV, $f_\pi = 0.092\,$GeV, $f_K=0.110\,$GeV \cite{Olive:2016xmw};
and $\mathpzc{w}_M^2 = e_{\bar q} \mathpzc{w}_{\bar q}^2(Q^2) + e_{u}\mathpzc{w}_u^2(Q^2)$,
%%  http://pdg.lbl.gov/2010/reviews/rpp2010-rev-pseudoscalar-meson-decay-cons.pdf
%%\begin{subequations}
%%\label{weightings}
%%\begin{align}
%\mathpzc{w}_M^2 & =& e_{\bar q} \mathpzc{w}_{\bar q}^2 + e_{u}\mathpzc{w}_u^2,  \\
%
%%\mathpzc{w}_{\bar q} &= \tfrac{1}{3}\int_0^1 dx\, \tfrac{1}{1-x}\,\varphi_M(x;Q^2) \,, \\
%%\mathpzc{w}_u  &= \tfrac{1}{3}\int_0^1 dx\, \tfrac{1}{x}\,\varphi_M(x;Q^2) \,,
%%\end{align}
%%\end{subequations}
\begin{equation}
\label{weightings}
\mathpzc{w}_{f} = \tfrac{1}{3}\int_0^1 dx\, \mathpzc{g}_f(x) \,\varphi_M(x;Q^2) \,,
\end{equation}
$\mathpzc{g}_u(x) = 1/x$, $\mathpzc{g}_{\bar q}(x) = 1/(1-x)$,  $e_{u}=2  e_{\bar q} = (2/3)$,  $\bar q = \bar s$ ($K^+$) or $\bar d$ ($\pi^+$) and $\varphi_M(x;Q^2)$ is the meson's scale-dependent leading-twist parton distribution amplitude (PDA).  The $\pi^0$ elastic form factor is identically zero owing to charge conjugation invariance; and a prediction for the neutral kaon is obtained via $e_u \to e_d = (-1/3)$.

The value of $Q_0$ is not predicted by perturbative QCD; but, fortunately, continuum bound-state methods have reached the point where $F_M(Q^2)$ can be calculated directly on the entire domain of spacelike momenta, thereby enabling $Q_0$ to be located.  This was accomplished for the pion in Ref.\,\cite{Chang:2013nia}.  Herein, we both refine the method and extend it to produce a wide range of verifiable form factor predictions, including statements, \emph{e.g}.\  about their behaviour at large timelike momenta.
%%strong-interaction theory ... collinear factorisation
%%spacelike vs. timelike \ldots experiment \ldots ratio
%%prediction IR to UV \ldots test collinear factorisation \ldots predict ratio in large domain \ldots project into timelike region \ldots meat and motivation for new experiments (spacelike and revisit timelike)

\smallskip

\noindent\emph{II.$\;$Computational Method}.\,---\,%
At leading order in the symmetry-preserving scheme for bound-state computations reviewed in Refs.\,\cite{Maris:2003vk, Chang:2011vu}, \emph{i.e}.\ the Dyson-Schwinger equation (DSE) rainbow-ladder (RL) truncation, kaon form factors can be computed as follows ($q=u$, $d$) \cite{Maris:2000sk}:
\begin{subequations}
\label{RLFKq}
\begin{align}
F_K(Q^2) & = e_q F_K^q(Q^2) + e_{\bar s} F_K^{\bar s}(Q^2)\,, \label{FKFuFs}\\
\nonumber
P_\mu F_K^q(Q^2) & = {\rm tr}_{\rm CD} %\int_{dk}^\Lambda
\int\! \frac{d^4 k}{(2\pi)^4}\,
\chi_\mu^q(k+p_o,k+p_i) \\
&  \quad \times \Gamma_K(k_i;p_i)\,S_s(k)\,\Gamma_K(k_o;-p_o)\,,  \label{RLAmp1}
\end{align}
\end{subequations}
with a similar expression for $F_K^{\bar s}(Q^2)$, where $Q$ is the in\-coming photon momentum, the trace is over colour and spinor indices, $p_{o,i} = P\pm Q/2$, $k_{o,i}=k+p_{o,i}/2$, $p_{o,i}^2 = -m_K^2$, $m_K$ is the kaon mass.  The calculation also requires quark propagators, $S_f$, $f=u(=d)$, $s$,
%%\begin{equation}
%%S_f(k) = -i \gamma\cdot k \sigma_V^f(k^2) + \sigma_S^f(k^2)\,,
%%\end{equation}
which, consistent with Eq.\,\eqref{RLAmp1}, are obtained from the rainbow-truncation gap equation; the kaon Bethe-Salpeter amplitude, $\Gamma_K$, computed in RL truncation; and consistent unamputated dressed-quark-photon vertices, $\chi_\mu^f$.

The leading-order result for $F_K(Q^2)$ is now determined once an interaction kernel is specified for the rainbow gap equation.  We use that explained in Ref.\,\cite{Qin:2011dd}, whose interaction strength is determined by a product: $D\omega = m_G^3$.  With $m_G$ fixed, results for properties of numerous ground-state hadrons are independent of the value of $\omega \in [0.4,0.6]\,$GeV \cite{Maris:2003vk, Eichmann:2008ae, Eichmann:2008ef, Qin:2011dd, Chang:2011vu, Qin:2011xq}: we use $\omega =0.5\,$GeV.  With this kernel, $f_\pi=0.092\,$GeV, $m_\pi=0.14\,$GeV and $f_K=0.11\,$GeV, $m_K=0.49\,$GeV are obtained with $m_G^{\zeta=2\,{\rm GeV}}=0.87\,$GeV and one-loop evolved current-quark masses $m_u^{\zeta=2{\rm GeV}}=4.7\,$MeV, $m_s^{\zeta=2{\rm GeV}}=112\,$MeV.

One may now evaluate the integrals in Eq.\,\eqref{RLFKq} using the algorithms introduced in Refs.\,\cite{Chang:2013pq, Chang:2013nia}.  Namely, the integrands are represented using the generalised Nakanishi interpolations of $S_{u,s}$ and $\Gamma_{K}$ described in Ref.\,\cite{Shi:2015esa}, of which the former also serve to express the unamputated photon-quark vertices, $\chi_\mu^q$ \cite{Chang:2013nia}.
With each element in Eq.\,\eqref{RLFKq} expressed via a generalised spectral representation, computation of $F_K(Q^2)$ reduces to the act of summing a series of terms, all of which involve a single four-momentum integral.  The integrand denominator in every term is a product of $k$-quadratic forms, each raised to some power.
Within each such term, one employs a Feynman parametrisation in order to combine the denominators into a single quadratic form, raised to the appropriate power.  A sensible change of variables then enables one to evaluate the four-momentum integration using standard algebraic methods.
After calculation of the four-momentum integration, evaluation of the individual term is complete after one computes a finite number of simple integrals; namely, integrations over Feynman parameters and the spectral integral.
The complete result for $F_K(Q^2)$ follows after summing the series.

\smallskip

\begin{figure}[t!]
\begin{centering}
%\vspace*{17.5em} %
\includegraphics[clip,width=0.75\linewidth]{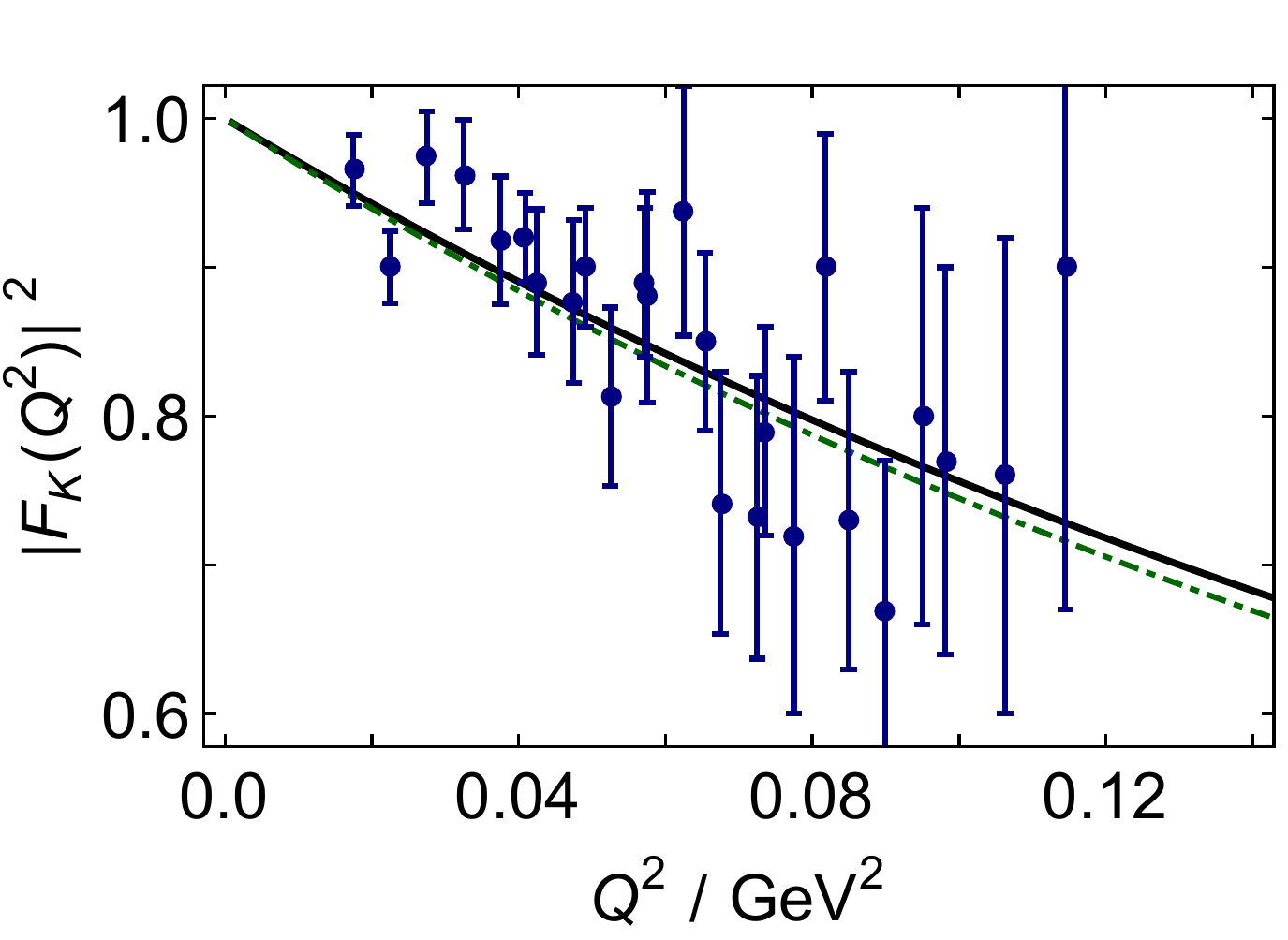}
\end{centering}
\caption{\label{Fig1}
 Solid curve -- our $K^+$ form factor; and dot-dash\-ed curve (green) -- result from Ref.\,\protect\cite{Maris:2000sk}.
%, whose boundary is marked by the vertical dotted line;
Data: Refs.\,\cite{Dally:1980dj, Amendolia:1986ui}.}
\end{figure}

\begin{figure}[t!]
\begin{centering}
%\vspace*{17.5em} %
\includegraphics[clip,width=0.75\linewidth]{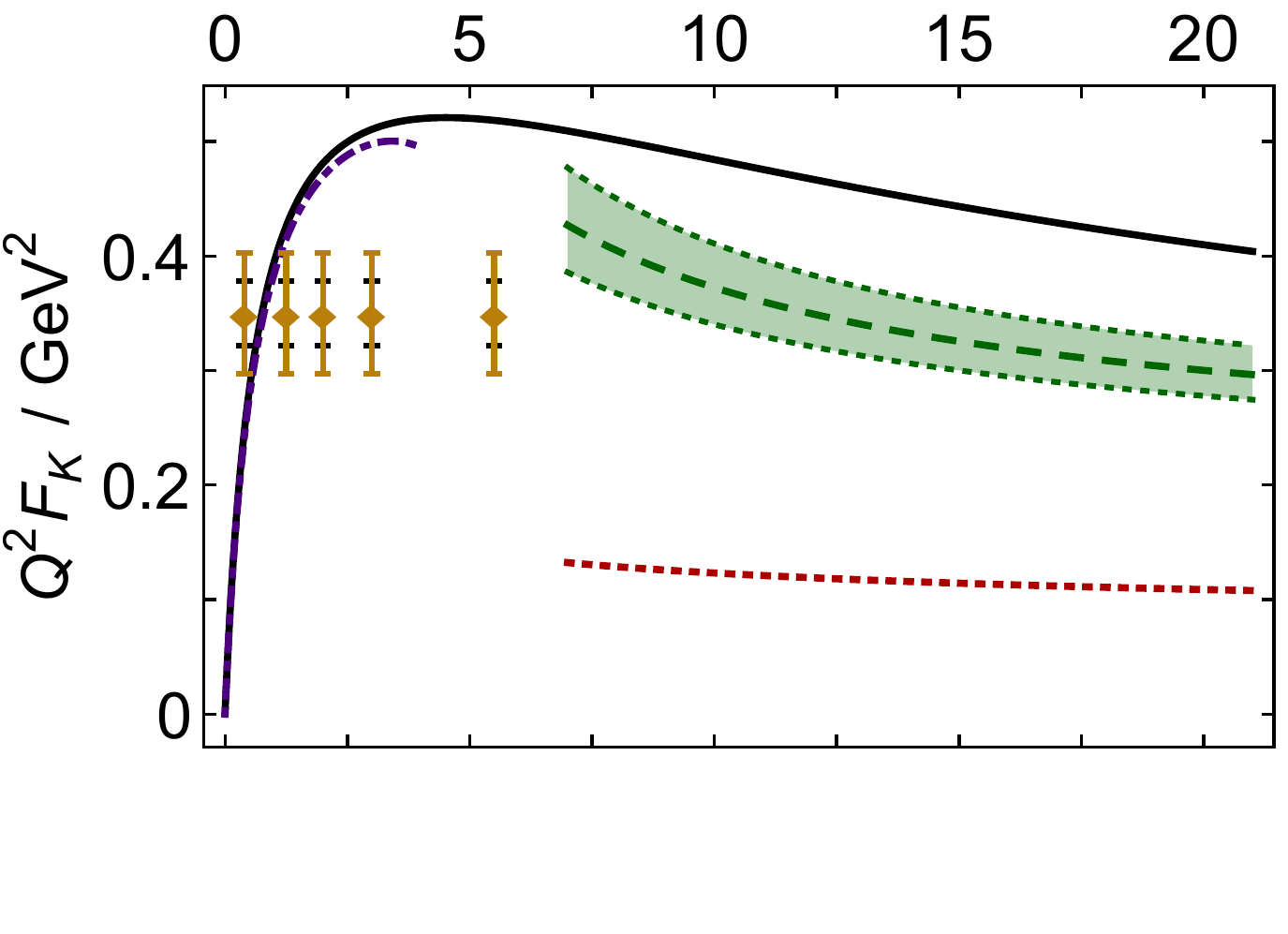}\vspace{-9.7ex}

\includegraphics[clip,width=0.75\linewidth]{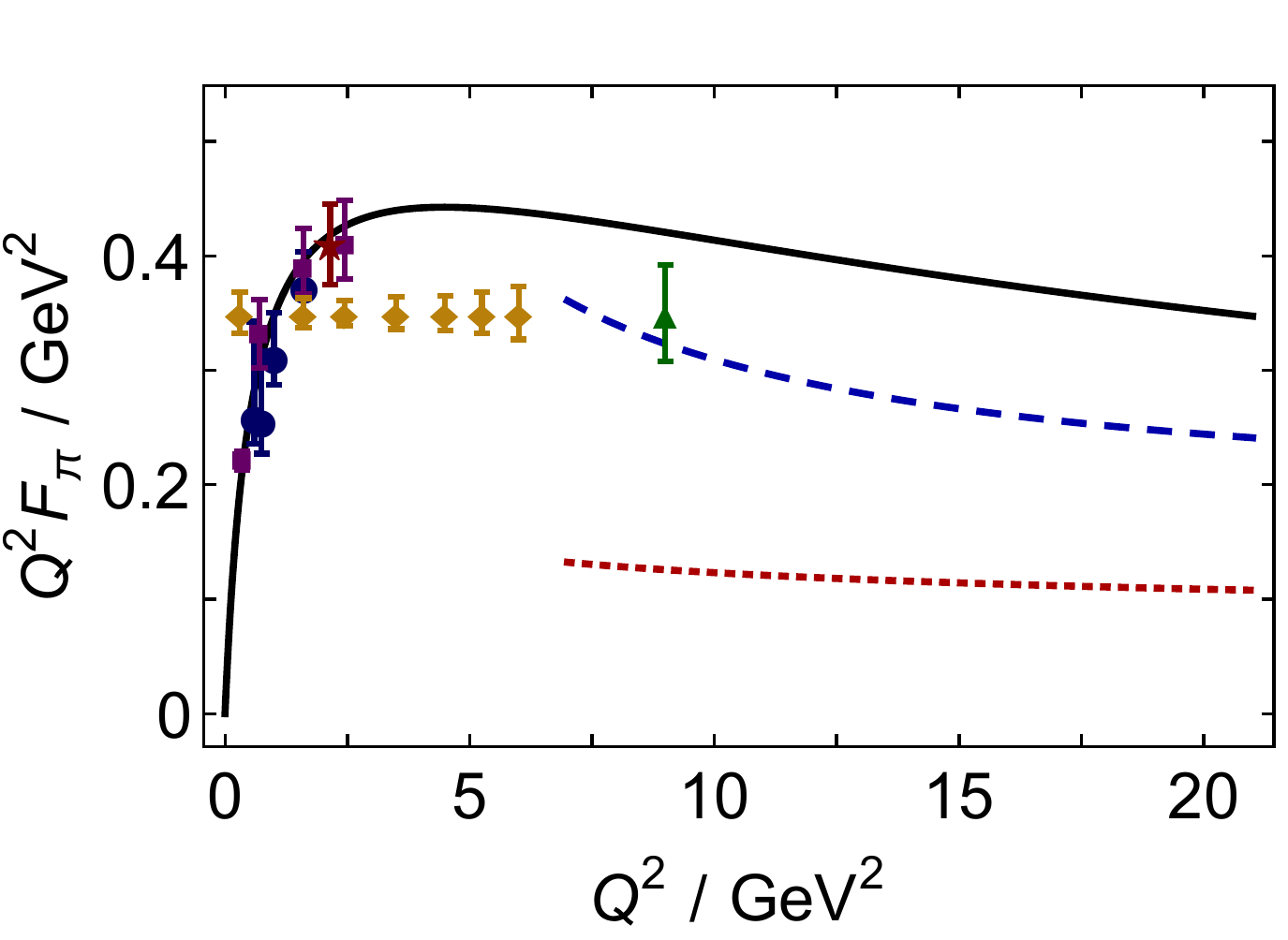}

\end{centering}
\caption{\label{Fig2}
\emph{Upper panel}.  Solid curve -- Eq.\,\eqref{RLFKq} prediction for $Q^2 F_K(Q^2)$;
dot-dashed curve (indigo) -- result from Ref.\,\protect\cite{Maris:2000sk}, which is limited to the domain $Q^2<4\,$GeV$^2$;
and dashed curve and band (green) -- result produced by the hard scattering formula, Eqs.\,\eqref{kaonUV}, \eqref{kaonPDA}.
Filled diamonds -- data anticipated from a forthcoming experiment \cite{E12-09-011}: the two error estimates differ in their assumptions about the $t$- and model-dependence of the form factor extractions \cite{Horn:2016rip}.
%%%
\emph{Lower panel}.  Solid curve -- prediction for $Q^2 F_\pi(Q^2)$;
and dashed curve (blue) -- result produced by the hard scattering formula, Eqs.\,\eqref{kaonUV}, \eqref{pionPDA}.
Data.  Star \cite{Horn:2007ug}, circles and squares \cite{Huber:2008id}; and diamonds and triangle, anticipated reach and accuracy of forthcoming experiments  \protect\cite{E1206101, E1207105}.
In both panels, the dotted curve (red) is Eq.\,\eqref{kaonUV} computed with the conformal limit PDA, $\varphi^{\rm cl}(x)=6 x(1-x)$.
}
\end{figure}

\noindent\emph{3:Results and Projections}.\,---\,The $K^+$ form factor, computed from Eq.\,\eqref{RLFKq} as described above,%
\footnote{For completeness, the $\chi_\mu^f$ include a parameter, $\eta_f$, which modulates the dressed-quark anomalous magnetic moment \cite{Singh:1985sg, Kochelev:1996pv, Bicudo:1998qb, Chang:2010hb}.  With $\eta_u=0.5$, $\eta_s=0.4$,  matching modern estimates \cite{Chang:2010hb, Bashir:2011dp, Qin:2013mta, Chang:2011tx}: $r_\pi=0.66\,$fm, $r_K=0.58\,$fm.  Setting $\eta_u=0=\eta_s$ reduces $r_{\pi,K}$ by $\lesssim 3$\% and has no visible impact on the curves drawn herein.  Hence, the impact of realistic dressed-quark anomalous magnetic moments on pseudoscalar meson form factors is small \cite{Chang:2013nia}.
}
is depicted in Figs.\,\ref{Fig1} and \ref{Fig2}.  The result is practically equivalent to that described in Ref.\,\cite{Maris:2000sk} on $Q^2 \lesssim 4\,$GeV$^2$, which is the entire domain accessible with the algorithms employed therein.  Here, however, we deliver a prediction for $F_K(Q^2)$ that extends to the entire domain of spacelike momenta; and this enables the first, realistic comparison with the prediction of Eq.\,\eqref{kaonUV}, so long as the kaon PDA is known.

A simultaneous computation of $\pi$ and $K$ PDAs is reported in Ref.\,\cite{Shi:2015esa}; and pointwise forms inferred from lattice-QCD computations of the distributions' low-order Mellin moments \cite{Arthur:2010xf, Braun:2015axa} are reported in Refs.\,\cite{Horn:2016rip, Segovia:2013eca}.  For the $\pi$, a clear picture has emerged \cite{Mikhailov:1986be, Brodsky:2006uqa, Chang:2013pq, Zhang:2017bzy}: $\varphi_\pi(x)$ is concave and markedly dilated compared to the conformal limit result, $\varphi^{\rm cl}(x)=6x(1-x)$, \emph{viz}.\ in RL truncation \cite{Chang:2013pq},
\begin{equation}
\label{pionPDA}
\varphi_\pi(x) = 1.77\,[x(1-x)]^{0.30}\,.
\end{equation}

The $K^+$ PDA has similar characteristics; but, in addition, it is skewed, so that the $\bar s$-quark carries more of the bound-state's light-front momentum.  However, the precise amount of skewing and dilation are unknown owing to disagreements between extant estimates.  One may only say \cite{Shi:2015esa, Horn:2016rip}: $\varphi_{K}$ is less dilated than $\varphi_{\pi}$; and the maximum of $\varphi_{K^+}$ lies in the neighbourhood of $x=0.56$.

A benefit of our simultaneous computation of $\pi$ and $K$ form factors (Fig.\,\ref{Fig2}) can be exploited here, \emph{viz}.\ using Eq.\,\eqref{kaonUV} and the method in Ref.\,\cite{Qin:2017lcd}, the above constraints can be employed to determine $\varphi_{K}$ from the computed value of $F_K(Q^2)/F_\pi(Q^2)$ on $Q^2 \simeq 8\,$GeV$^2$, with the result
\begin{equation}
\label{kaonPDA}
\varphi_K(x) = \mathpzc{n}_{\,\alpha\beta}\, x^\alpha (1-x)^\beta\!,\;
\alpha=0.39(4) \,, \; \beta = 0.31(4)\,,
\end{equation}
$\mathpzc{n}_{\,\alpha\beta}=\Gamma (\alpha +\beta +2)/[\Gamma (\alpha +1) \Gamma (\beta +1)]$.
% is a constant that ensures the zeroth moment of this PDA is unity.
%
%The analyses in Refs.\,\cite{Chang:2013nia, Qin:2017xxx} establish the validity of this approach on $Q^2 \gtrsim 8\,$GeV$^2$.
We have introduced an error to express uncertainty in the dilation, measured by $\langle (2x-1)^2\rangle_K = 0.271(5)$.  The size of the error is chosen to match that in modern lattice-QCD estimates for this moment of the pion's PDA \cite{Braun:2015axa}.  The associated result $\langle (2x-1)\rangle_K=0.0296(9)$ has greater uncertainty but far lesser impact \cite{Horn:2016rip}.

Using Eq.\,\eqref{kaonPDA} in Eq.\,\eqref{kaonUV} yields the (green) dashed curve and band in the upper panel of Fig.\,\ref{Fig2}.  Like the (blue) dashed curve in the lower panel, it is a near match in magnitude to our complete prediction, but far above the (red) dotted curve, obtained from Eq.\,\eqref{kaonPDA} by using $\mathpzc{w}_{\bar q}=\mathpzc{w}_{u}=1$, \emph{i.e}.\ frozen at their conformal-limit values.
Crucially, too, its evolution matches that of the RL prediction on $Q^2 \gtrsim 12\,$GeV$^2$.
Contrasting this with the evolution of the frozen-PDA conformal-limit result leads us to describe a qualitative improvement of over Ref.\,\cite{Chang:2013nia}.

It has long been known \cite{Lepage:1980fj} that, whilst producing the right $1/Q^2$ behaviour, symmetry-preserving computations via Eq.\,\eqref{RLFKq} (or its analogues for related processes) typically fail to generate the correct anomalous dimension and therefore yield form factors with wrong-power logarithmic scaling violations.
This can be understood by noting that the meson's wave function must evolve with resolving scale just as its leading-twist PDA so that the dressed-quark and -antiquark degrees-of-freedom, in terms of which the wave function is expressed at a given scale $Q^2$, can split into less-well-dressed partons via the addition of gluons and sea quarks as prescribed by QCD dynamics.  Such effects are incorporated in bound-state problems when the complete quark-antiquark scattering kernel is used; but aspects are lost when that kernel is truncated, and so it is with the RL truncation.
As emphasised in recent studies of neutral pseudoscalar meson transition form factors \cite{Raya:2015gva, Raya:2016yuj}, this is a critical flaw now that one can use QCD-connected input to make predictions at arbitrarily large $Q^2$ because it precludes any valid attempt to match theory with experiment.  Recognising that, Refs.\,\cite{Raya:2015gva, Raya:2016yuj} introduced a remedy; and the supplemental material explains how we adapt and employ that method herein for elastic form factors.

\begin{figure}[t!]
\begin{centering}
%\vspace*{17.5em} %
\includegraphics[clip,width=0.75\linewidth]{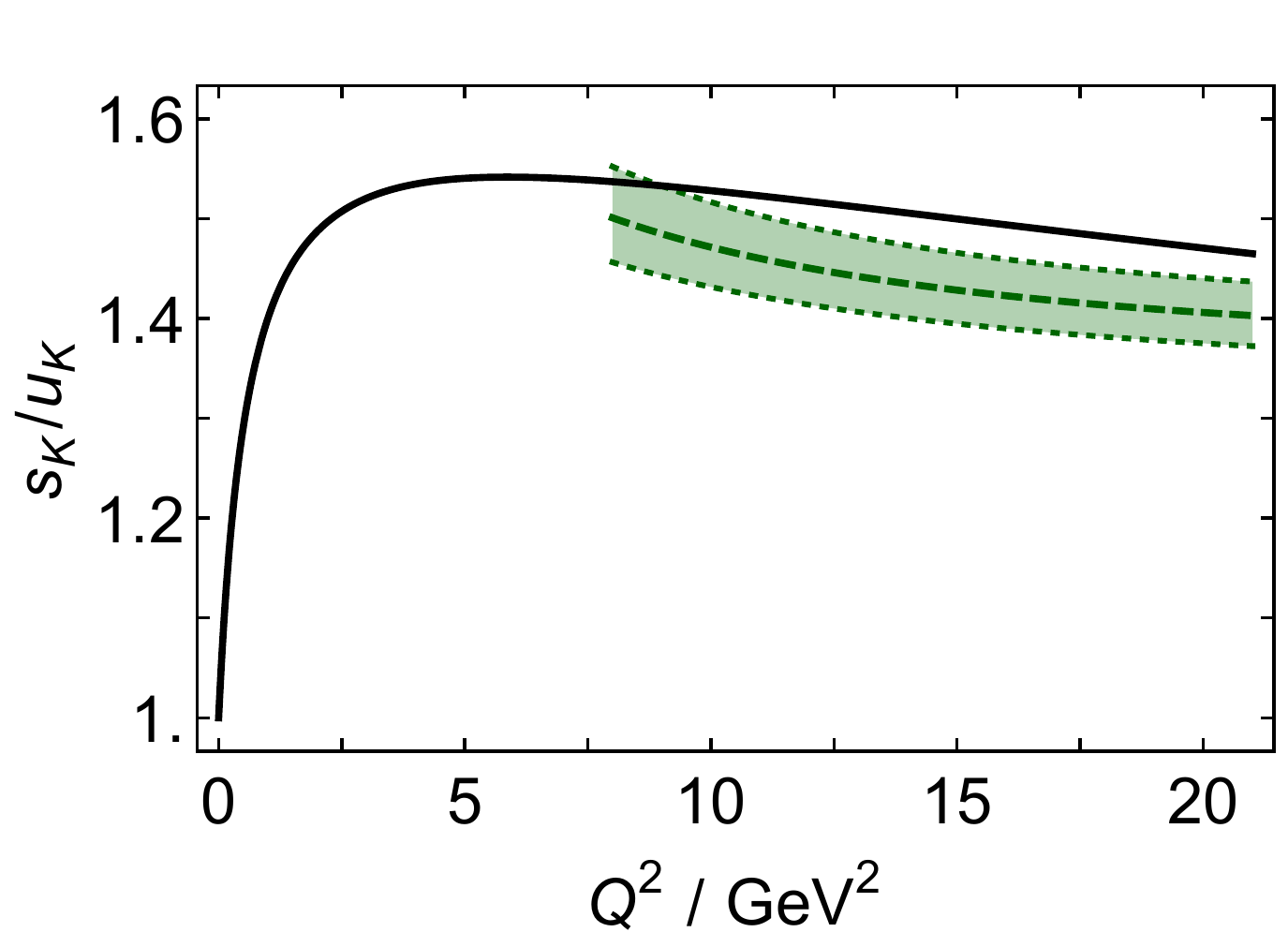}
\end{centering}
\caption{\label{Fig3}
Solid curve -- ratio of $\bar s$- and $u$-quark contributions to the $K^+$ form factor; and dashed curve and band (green) -- prediction of the hard-scattering formula, Eqs.\,\eqref{kaonUV}, \eqref{kaonPDA}.
}
\end{figure}

%% ... I have run Chen's code.  Incorporating the kaon's pseudovector component, the ratio evolves to unity.  Omitting the pseudovector component, as idiots do, the ratio continues to grow without limit.
%% ... Leading-twist behaviour of meson form factor .. hard scattering ... involves one-gluon exhange ... generates a pseudsovector component in bound-state ... that is omitted in triangle diagram unless meson already contains the pseudovector term, which is then generated self-consistently.
A flavour-separation of the $K^+$ form factor is depicted in Fig.\,\ref{Fig3}.  With $s_K=F_K^{\bar s}$, $u_K=F_K^u$ from Eq.\,\eqref{FKFuFs}, current conservation ensures $s_K/u_K = 1$ at $Q^2=0$.
This ratio must increase on some domain of $Q^2>0$ because the effective mass of a dressed $s$-quark is greater than that of a dressed $u$-quark or, equally, the lightest vector meson that can couple to a $\bar s \gamma_\mu s$ current ($\phi$) is heavier than that which can couple to a $\bar u \gamma_\mu u $ current ($\rho$) \cite{Maris:2000sk, Chen:2012txa}.
Notwithstanding that, all analyses which faithfully preserve the structure of pseudoscalar meson bound-state amplitudes generated by a vector$\,\otimes\,$vector interaction \cite{Maris:1997tm, Maris:1998hc, GutierrezGuerrero:2010md, Roberts:2010rn, Roberts:2011wy, Chen:2012txa, Serna:2016kdb, Bedolla:2016yxq}, thereby ensuring Eq.\,\eqref{RLFKq} yields internally-consistent leading-twist power-law behaviour, produce a ratio $s_K/u_K$ that reaches a maximum at some nonzero value of $Q^2$.  Thereafter, $s_K/u_K\to 1^+$.  The height and location of the maximum are a measure of dynamics, and we predict a peak value $s_K/u_K\approx 1.5$ at $Q^2\approx 6\,$GeV$^2$.
%
%Naturally, the maximum must lie below the lower boundary of the domain upon which Eq.\,\eqref{kaonUV} is valid because the ratio is monotonically decreasing
Given that Eq.\,\eqref{kaonUV} provides a semiquantitatively accurate description of the $K^+$ form factor on $Q^2\gtrsim 8\,$GeV$^2$ (Fig.\,\ref{Fig2}), then thereupon one should also obtain a reliable estimate of $s_K/u_K$ using the elements of the hard scattering formula.  This is evidently the case.
%% Contact interaction ... peak value 1.2 at around 1GeV^2

\begin{figure}[t!]
\begin{centering}
%\vspace*{17.5em} %
\includegraphics[clip,width=0.75\linewidth]{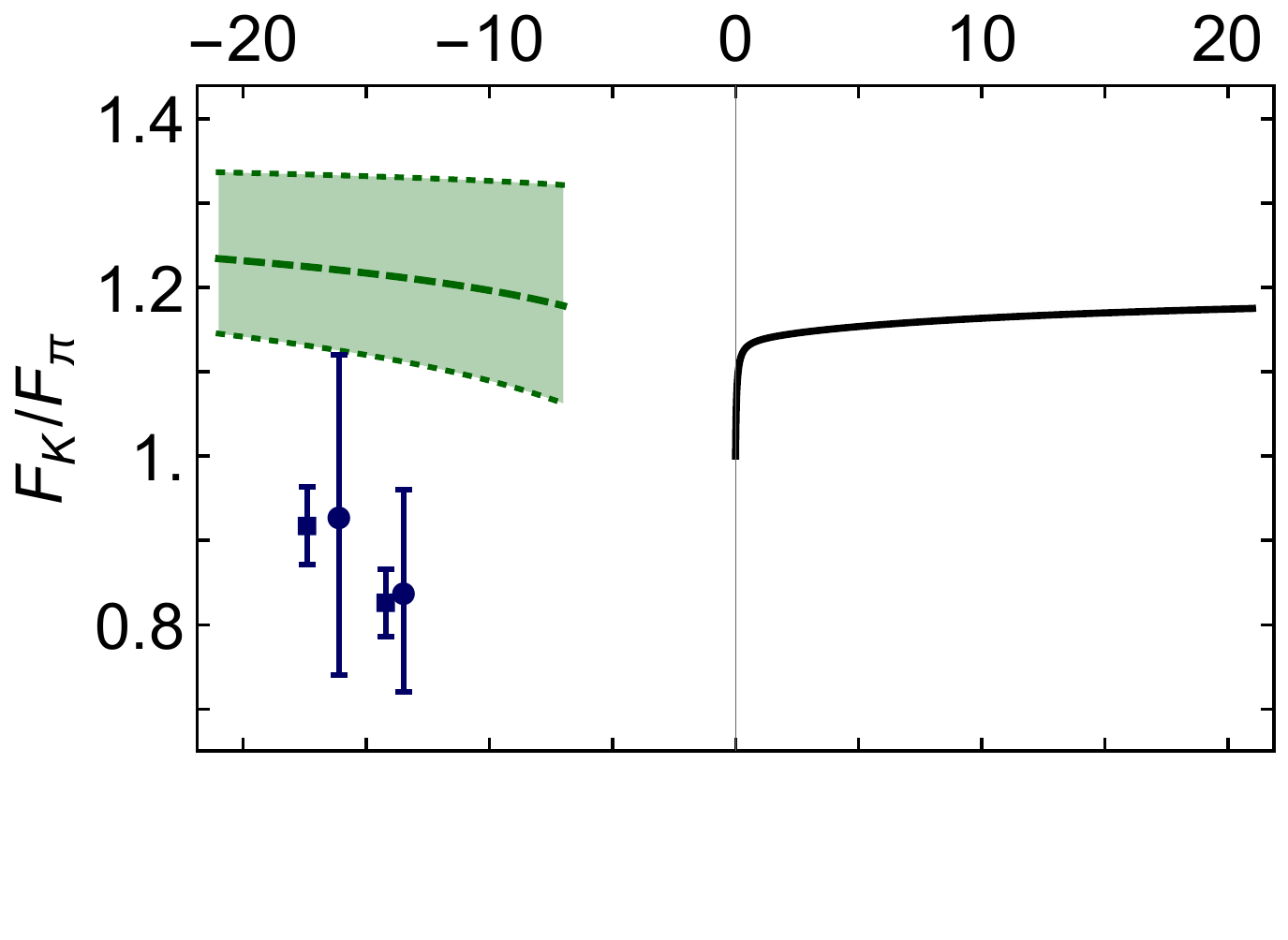}\vspace*{-9.5ex}

\includegraphics[clip,width=0.78\linewidth]{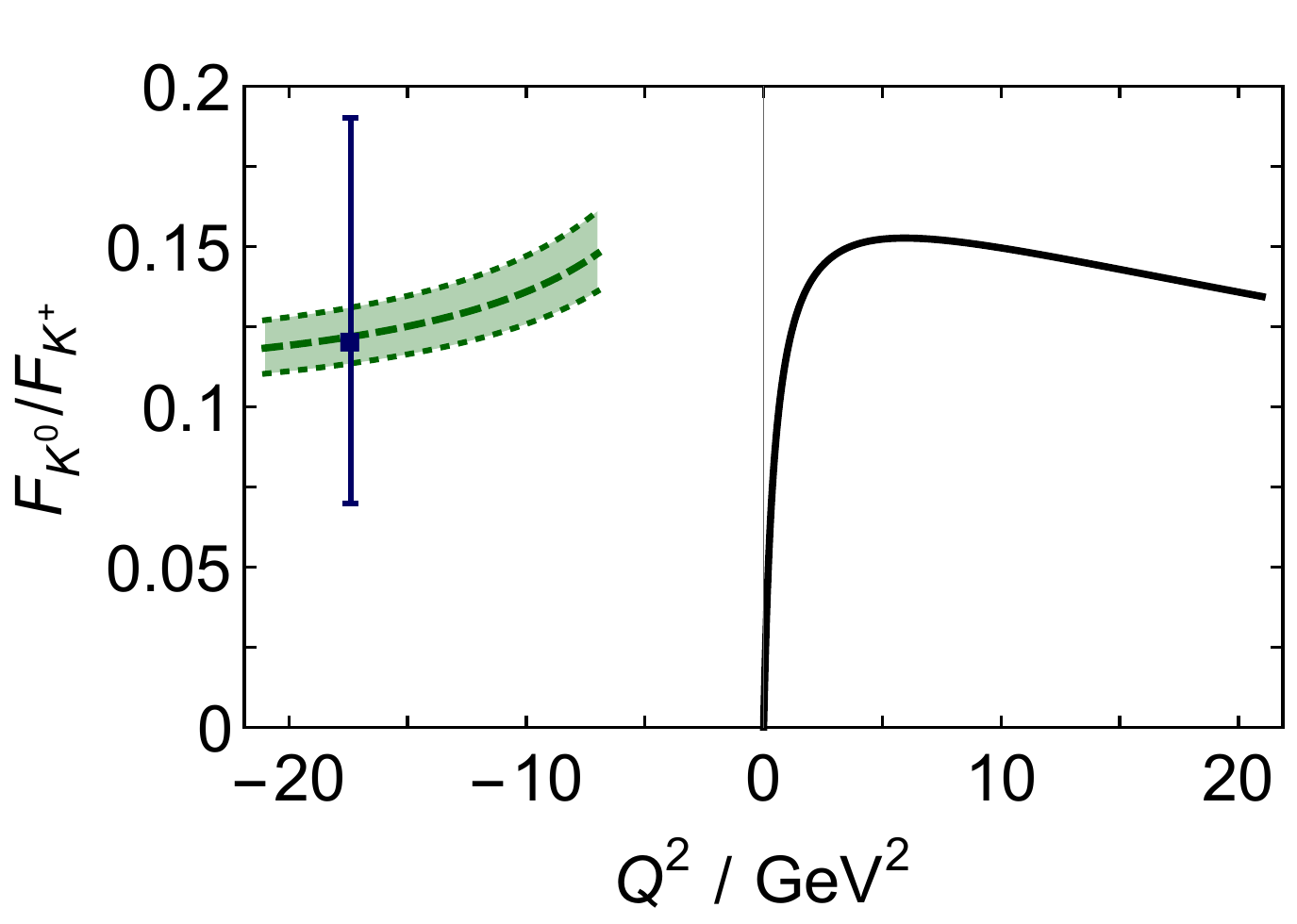}\hspace*{1ex}
\end{centering}
\caption{\label{Fig4}
\emph{Upper panel}.  Charged kaon-to-pion ratio.  Solid curve at spacelike momenta -- direct calculation, Eq.\,\eqref{RLFKq}.  Dashed curve and band (green) -- prediction for the timelike behaviour, derived from Eqs.\,\eqref{kaonUV}, \eqref{pionPDA}, \eqref{kaonPDA}.  Data from Ref.\,\cite{Seth:2012nn}.
\emph{Lower panel}.  Ratio of neutral-to-charged kaon form factors.  Solid curve -- direct calculation.  We compute $r_{K^0}^2 = -(0.21\,{\rm fm})^2$ \emph{cf}.\ experiment \cite{Molzon:1978py}: $r_{K^0}^2 = -(0.24 \pm 0.08\,{\rm fm})^2$.
Dashed curve and band (green) -- prediction for the timelike behaviour of this ratio derived from Eqs.\,\eqref{kaonUV}, \eqref{kaonPDA}.  Datum from Ref.\,\cite{Seth:2013eaa}: the error bar marks the 90\% confidence interval.
}
\end{figure}

We depict form factor ratios in Fig.\,\ref{Fig4}.
%: charged kaon-to-pion in the upper panel; and neutral-to-charged kaon in the lower panel.
In accordance with Eq.\,\eqref{kaonUV}, our calculated result for $F_K/F_\pi$ rises logarithmically to $f_K^2/f_\pi^2\approx 1.42$ as $Q^2\to \infty$, whereas $F_{K^0}/F_{K^+}$ vanishes.  As was to be anticipated from Fig.\,\ref{Fig2}, we predict that these conformal limit values are inaccessible at terrestrial facilities.  On the other hand, used with the PDAs appropriate to the probe scale, the hard scattering formulae are applicable.  We therefore capitalise on the fact that spacelike and timelike form factors are identical at leading order in $\alpha_s$, and use Eqs.\,\eqref{kaonUV}, \eqref{pionPDA}, \eqref{kaonPDA} to make projections for these ratios at timelike momenta beyond the resonance region, \emph{i.e}. on $t=-Q^2 \gtrsim 8\,$GeV$^2$.
Evidently, the prediction for $F_{K^0}/F_{K^{+}}$ at $t=17.4\,$GeV$^2$ obtained in this way is consistent with the only existing measurement and explains it as the natural outcome of using PDAs appropriate to the scale of the experiment.

The situation is less clear for $F_{K}/F_{\pi^{+}}$: we predict $F_{K}/F_{\pi^{+}}>1$ on $t>8\,$GeV$^2$, whereas extant data lie below unity \cite{Seth:2012nn}.  We can identify no reasonable means by which our direct computation of this ratio at spacelike momenta could be less-than one at any value of $Q^2$: charge conservation means $F_{K}/F_{\pi^{+}}=1$ at $Q^2=0$, the ordering of charge radii ensures it rises as $Q^2$ increases, and the absence of another set of mass-scales suggests that the conformal limit ($f_K^2/f_\pi^2\approx 1.42$) should be approached monotonically from below.  These features are expressed in the semi-quantitative agreement between the hard-scattering formulae and our direct calculations (Figs.\,\ref{Fig2}, \ref{Fig3}), and support the soundness of the timelike prediction based on Eqs.\,\eqref{kaonUV}, \eqref{pionPDA}, \eqref{kaonPDA}.  Studying the separate empirical results for the $\pi$ and $K$ form factors at timelike momenta \cite{Seth:2012nn}, one might question their normalisations because, mapped simply to spacelike momenta and compared with our direct calculations, the $\pi$ measurements are a factor of $\sim 2$ larger, and those for the $K$, greater by a factor of $\sim 1.5$.  Notably, a mismatch of relative normalisations would cancel in $F_{K^0}/F_{K^{+}}$.

\smallskip

\noindent\emph{4:Concluding Remarks}.\,---\,Using a single bound-state interaction kernel, fully determined by just one parameter, we presented a unified description of $\pi$ and $K$ elastic form factors.
%
%In doing so, we arrived at a refined prediction for the kaon's twist-two parton distribution amplitude (PDA); and t
This enabled us to show that, when used with PDAs computed at the probe scale, which express dynamical consequences of emergent phenomena within the Standard Model (SM), leading-order, leading-twist hard-scattering formulae derived for $\pi$ and $K$ elastic form factors are both accurate to 25\% on $Q^2 \simeq 8\,$GeV$^2$, becoming more reliable as $\ln Q^2$ is increased.
Our analysis also yields projections for the separate $\bar s$- and $u$-quark contributions to the $K^+$ form factor. Eliminating the quark-charge weight-factors, the ratio of these contributions is unity at $Q^2=0$, increases monotonically to a peak value of roughly 1.5 at $Q^2\approx 6\,$GeV$^2$, and thereafter returns logarithmically to unity, again in agreement with the SM.
With continuing developments in the numerical simulation of lattice-regularised quantum chromodynamics, it should be possible to validate this unified body of predictions in the foreseeable future \cite{Koponen:2017fvm, Chambers:2017tuf}.

Having established the domain of reliability for the hard scattering formulae, we argued that they may be used to make predictions for $\pi$ and $K$ form factors at timelike momenta beyond the resonance region.  Some data exist on this domain, but a comparison between experiment and our predictions is currently inconclusive.  Notwithstanding that, the prospects for improving measurements in the timelike region are excellent given the capabilities of existing and planned $e^+ e^-$ colliders \cite{Holt:2012gg}.

Our study reveals noticeable differences between the distribution of strange and normal matter within the strong interaction's pseudo-Nambu-Goldstone modes.  Consequently, they should serve to spur and guide new experiments at both spacelike and timelike momenta.

\smallskip

\noindent\textbf{Acknowledgments}.
%\section*{Acknowledgments}
%
We are grateful to R.~Ent, T.~Horn and C.~Mezrag for insightful comments.
This research was facilitated by the
\emph{2$^{\it nd}$ Sino-Americas Workshop and School on the Bound-State Problem in Continuum QCD}, Central China Normal University, Wuhan, China, and the workshop on \emph{Novel theory for new facilities in hadron physics}, Nankai University, Tianjin, China.
Research supported by:
the National Natural Science Foundation of China (contract nos.\ 11435001 and 11175004);
the National Key Basic Research Program of China (contract nos.\ G2013CB834400 and 2015CB85690);
the Chinese Ministry of Education, under the \emph{International Distinguished Professor} programme;
U.S.\ Department of Energy, Office of Science, Office of Nuclear Physics, under contract no.~DE-AC02-06CH11357;
and National Science Foundation, under grant no.\ NSF-PHY1516138.
%

%--
%\bibliographystyle{cj} %%%-- Correct for FBS
%\bibliography{../../../CollectedBiB}
%-
%%\bibliographystyle{../../../../zProc/z10/z10KITPC/h-physrev4}
%%\bibliography{../../../../CollectedBiB}

\medskip

\centerline{\rule{10em}{0.1ex}}

\medskip

\setcounter{equation}{0}
\renewcommand{\theequation}{A\arabic{equation}}
\renewcommand{\thetable}{A\arabic{table}}

\noindent\emph{Supplemental material}.\,---\,Consider Eq.\,\eqref{RLFKq}.  If one uses $S_f(k)=1/[i\gamma\cdot k + M_f]$, where the dressed-quark masses are constant, with $M_s \approx 1.2 M_u$, then symmetries ensure $\chi_\mu^f = S_f(k_{\rm out}) \gamma_\mu S_f(k_{\rm in})$ is an adequate representation of the photon-quark vertex.
The remaining element is the kaon Bethe-Salpeter amplitude.  Following Refs.\,\cite{Maris:1997tm, Maris:1998hc, Chang:2013pq, Chen:2016sno}, one learns that realistic outcomes are ensured by:
\begin{align}
\nonumber
\Gamma_K(k;&P)  =
\alpha_{K}\frac{\Lambda_{K}}{f_{K}}\Bigg\{\int_{-1}^{1} dz \frac{\Lambda_{K}^{2}}{(k+\frac{z}{2}P)^{2}+\Lambda_{K}^{2}}  \\
\nonumber
& \times \bigg[i\gamma_{5}(1+\varepsilon z)\rho_{\nu_{E}}(z)  \\
& + g_{F}\gamma_{5} (\gamma\cdot P-4 (\tfrac{3}{2}+\nu_{F})z\gamma\cdot k)\rho_{\nu_{F}}(z) \bigg]
\Bigg\}\,, \label{kaonmodelBSA}
\end{align}
with
\begin{equation}
\rho_{\nu}(z)=\frac{\Gamma(\frac{3}{2}+\nu)}{\sqrt{\pi}\Gamma(1+\nu)}(1-z^{2})^{\nu}\,.
\end{equation}
The $\gamma_{5}$ term sets the scale of low-momentum observables; the $\gamma_5\gamma\cdot P$, $\gamma_5\gamma\cdot k$ contributions are necessary to ensure the correct form-factor power-law behaviour at large momentum transfers, and they are combined with relative-weight $[-4 (\tfrac{3}{2}+\nu_{F})]$ so as to eliminate a renormalisable divergence from the integral that defines the kaon's leptonic decay constant, $f_K$.

With these structures in hand, one can also evaluate the kaon's leading-twist PDA:
\begin{equation}
\label{kaonmodelPDA}
f_K \, \varphi(x;Q^2) = {\rm tr}_{\rm CD} \int_{dk}^{Q^2}\delta_n^u(k_\eta)\gamma_5\gamma\cdot n \,  \chi_K(k_\eta,k_{\bar\eta})\,,
\end{equation}
where
$\int_{dk}^{Q^2}$ is a Poincar\'e-invariant regularisation of the four-dimensional integral, with $Q^2$ setting the PDA's scale;
$\delta_n^u(k_\eta)=\delta(n\cdot k_\eta - x n\cdot P)$, $n^2=0$, $n\cdot P = -m_K$;
\begin{equation}
\chi_K(k_\eta,k_{\bar\eta}) = S_u(k_\eta) \Gamma_P(k_{\eta\bar\eta};P) S_s(k_{\bar \eta})\,,
\end{equation}
$k_{\eta\bar\eta} = [k_\eta+k_{\bar\eta}]/2$, $k_\eta = k + \eta P$, $k_{\bar\eta} = k - (1-\eta) P$, $\eta\in [0,1]$.  Inserting Eq.\,\eqref{kaonmodelBSA} in Eq.\,\eqref{kaonmodelPDA}, one arrives at an algebraic result:
\begin{equation}
\label{phiQ2}
\varphi(x;Q^2) = \mathpzc{n}_{\,\alpha\beta} \, x^\alpha (1-x)^\beta\,,
\end{equation}
where the $Q^2$-dependent exponents $\alpha$, $\beta$ are fixed by the values of $\nu_E$ and $\varepsilon$ in Eq.\,\eqref{kaonmodelBSA}.

Now, setting $m_K=0.49\,$GeV and $Q^2=4\,$GeV$^2=: Q_2^2$, one obtains unit charge, $f_K=0.11\,$GeV, and the PDA in Eq.\,\eqref{kaonPDA} with
$M_u=0.4\,$GeV,
$\Lambda_K=1.1 M_u$,
$\nu_E=-0.78$,
$\nu_F=1$,
$g_F=0.1\,{\rm GeV}^{-1}$,
$\varepsilon=0.10$,
$\alpha_K=1.48$.

It is now straightforward to chart the impact on the kaon form factor of leading-order QCD evolution \cite{Lepage:1979zb, Lepage:1980fj, Efremov:1979qk}.
One evolves the PDA from $Q_2^2$ to some new value, $Q^2$.  The evolved PDA may still be expressed using Eq.\,\eqref{phiQ2} and can therefore be recovered using Eqs.\,\eqref{kaonmodelBSA}, \eqref{kaonmodelPDA} so long as evolved values of $\nu_E$ and $\varepsilon$ are used.  In fact, on the domain $Q_2^2 < Q^2 \leq 40\,$GeV$^2$ one may use $\varepsilon = \,$constant; and the impact of evolution is accurately incorporated simply by using $(\zeta^2= Q^2-Q_2^2)$:
\begin{equation}
\label{nuEzeta}
\nu_E(\zeta^2) = -\frac{0.78 + 0.059 \zeta^2}{1 + 0.098 \zeta^2}\,.
\end{equation}

Thus informed, one evaluates the $\bar s$- and $u$-quark contributions to the $K^+$ form factor, defined by Eq.\,\eqref{RLFKq}, on $\zeta^2>0$, using the simple propagators and vertices described above and $\nu_E(\zeta^2)$ from Eq.\,\eqref{nuEzeta} in the expression for the Bethe-Salpeter amplitude, Eq.\,\eqref{kaonmodelBSA}.  Normalising to the values at $\zeta^2=0$, one obtains the following evolution functions for these separate contributions:
\begin{subequations}
\begin{align}
\mathpzc{E}_{\,u_K}(\zeta^2>0) & = \frac{1 + 0.10 \zeta^2 + 0.0011 \zeta^4}
    {1 + 0.12 \zeta^2 + 0.0019 \zeta^4}\,, \\
\mathpzc{E}_{\,s_K}(\zeta^2>0) & = \frac{1 + 0.033 \zeta^2 - 0.000011 \zeta^4 }
    {1 + 0.049 \zeta^2 +  0.000044 \zeta^4}\,.
\end{align}
\end{subequations}
Our final results for the meson form factors are obtained by incorporating these evolution factors into the individual pieces of Eq.\,\eqref{FKFuFs}:
\begin{align}
\nonumber
F_K^f(&Q^2)  = \hat F_K^f(Q^2)\\
& \times  [\theta(Q_2^2-Q^2) + \theta(Q^2-Q_2^2) \mathpzc{E}_{\,f_K}(Q^2-Q_2^2)]\,,
\end{align}
where $ \hat F_K^f(Q^2)$ is the result obtained directly from Eq.\,\eqref{RLAmp1}.  At $Q^2=20\,$GeV$^2$, each of these functions introduces a $\sim 15$\% suppression.

The procedure described in this supplement \cite{Raya:2015gva, Raya:2016yuj} assumes that the dressed-quark degrees-of-freedom defined by a RL computation renormalised at $Q_2^2=4\,$GeV$^2$ capture all relevant dynamics below that scale, an assumption supported by comparisons with extant data, and thereafter enables those degrees-of-freedom to evolve as prescribed by QCD.  As Figs.\,\ref{Fig2}, \ref{Fig3} demonstrate, the method repairs a known failing \cite{Lepage:1980fj} of symmetry-preserving computations of meson form factors via Eq.\,\eqref{RLFKq}, \emph{viz}.\ it solves the problem of wrong-power logarithmic scaling violations.  The method is readily simplified to suit the pion.
\end{document}